\documentclass[sigconf]{acmart}
\usepackage{caption}
\usepackage{graphicx}
\usepackage{xspace}
\usepackage{cleveref}
\usepackage{lipsum}
\usepackage[normalem]{ulem}
\usepackage{enumitem}% http://ctan.org/pkg/enumitem
\usepackage{geometry}
\usepackage{amsmath}
\setlist[itemize]{noitemsep, topsep=0pt}
\usepackage[utf8]{inputenc}
\usepackage[T1]{fontenc}
\usepackage{adjustbox}
\usepackage{subfigure}
\usepackage{multirow}
\usepackage[normalem]{ulem}

\usepackage{soul}
\usepackage{hyperref}
\hypersetup{
    colorlinks=true,
    linkcolor=blue,
    filecolor=magenta,
    urlcolor=cyan,
}

\usepackage{siunitx}
\usepackage{tikz}

\makeatletter
\DeclareRobustCommand\onedot{\futurelet\@let@token\@onedot}
\def\@onedot{\ifx\@let@token.\else.\null\fi\xspace}

\makeatother

\AtBeginDocument{%
  \providecommand\BibTeX{{%
    \normalfont B\kern-0.5em{\scshape i\kern-0.25em b}\kern-0.8em\TeX}}}

\begin{document}
% \title[]{\name: Detecting Counterfeit Powdered Food Products using a Single Commodity Smartphone} 
\title[]{Verifiable Dropout: Turning Randomness into a Verifiable Claim}

% % The "author" command and its associated commands are used to define
% % the authors and their affiliations.
% % Of note is the shared affiliation of the first two authors, and the
% % "authornote" and "authornotemark" commands
% % used to denote shared contribution to the research.

% \author{Kichang Lee, Sungmin Lee, Jaeho Jin, JeongGil Ko}
% \email{{kichang.lee, i.am.sungmin.lee,  jaeho.jin.eis, jeonggil.ko}@yonsei.ac.kr}
% \affiliation{%
%   \institution{Yonsei University}
%   \country{}
% }
\author{Kichang Lee}
\email{kichang.lee@yonsei.ac.kr}
\affiliation{%
  \institution{Yonsei University}
  \country{}
}
\author{Jaeho Jin}
\email{jaeho.jin.eis@yonsei.ac.kr}
\affiliation{%
  \institution{Yonsei University}
  \country{}
}
\author{Sungmin Lee}
\email{i.am.sungmin.lee@yonsei.ac.kr}
\affiliation{%
  \institution{Yonsei University}
  \country{}
}
\author{JeongGil Ko}
\email{jeonggil.ko@yonsei.ac.kr}
\affiliation{%
  \institution{Yonsei University}
  \country{}
}
%%%%%%%%%%%%%%%%%%%%%%%%%%%%%%%%%%%%%%%%%%
\begin{abstract}
Modern cloud-based AI training relies on extensive telemetry and logs to ensure accountability. While these audit trails enable retrospective inspection, they struggle to address the inherent non-determinism of deep learning. Stochastic operations, such as dropout, create an ambiguity surface where attackers can mask malicious manipulations as natural random variance, granting them plausible deniability. Consequently, existing logging mechanisms cannot verify whether stochastic values were generated and applied honestly without exposing sensitive training data. To close this integrity gap, we introduce \textit{Verifiable Dropout}, a privacy-preserving mechanism based on zero-knowledge proofs. We treat stochasticity not as an excuse but as a verifiable claim. Our approach binds dropout masks to a deterministic, cryptographically verifiable seed and proves the correct execution of the dropout operation. This design enables users to audit the integrity of stochastic training steps post-hoc, ensuring that randomness was neither biased nor cherry-picked, while strictly preserving the confidentiality of the model and data.
\end{abstract}
%%%%%%%%%%%%%%%%%%%%%%%%%%%%%%%%%%%%%%%%%%
\begin{comment}
\begin{CCSXML}
<ccs2012>
   <concept>
       <concept_id>10010520.10010553</concept_id>
       <concept_desc>Computer systems organization~Embedded and cyber-physical systems</concept_desc>
       <concept_significance>500</concept_significance>
       </concept>
   <concept>
       <concept_id>10010147.10010257.10010321</concept_id>
       <concept_desc>Computing methodologies~Machine learning algorithms</concept_desc>
       <concept_significance>500</concept_significance>
       </concept>
 </ccs2012>
\end{CCSXML}

\ccsdesc[500]{Computer systems organization~Embedded and cyber-physical systems}
\ccsdesc[500]{Computing methodologies~Machine learning algorithms}

\keywords{Federated learning, Mobile systems, Mobile AI}
\end{comment}
%%%%%%%%%%%%%%%%%%%%%%%%%%%%%%%%%%%
\settopmatter{printfolios=true} %page numbering
\settopmatter{printacmref=false} % Removes citation information below abstract
\renewcommand\footnotetextcopyrightpermission[1]{} % removes footnote with conference information in first column
%%%%%%%%%%%%%%%%%%%%%%%%%%%%%%%%%%%
\maketitle
%%%%%%%%%%%%%%%%%%%%%%%%%%%%%%%%%%%
%%%%%%%%%%%%%%%%%%%%%%%%%%%%%%%%%%%%%%%%%%%%%%%%%%%%%%%%%%%%%%%%%%%%%%%%%%%%%%%%%%%%

\section{Introduction}
\label{sec:intro}
Modern AI training is increasingly mediated by cloud platforms and managed ``training-as-a-service'' pipelines. As model and dataset scales grow~\cite{lee2025towards, lee2025gmt}, practitioners often outsource training to complex supply chains spanning hosted accelerators, distributed schedulers, containerized ML stacks, and third-party observability and artifact management. A practical advantage of this workflow is traceability: platforms retain execution logs that enable retrospective inspection and debugging. This has motivated a common narrative that cloud training can be made accountable through logging, monitoring, and post hoc verification.

\begin{figure}
    \centering
    \includegraphics[width=1\linewidth]{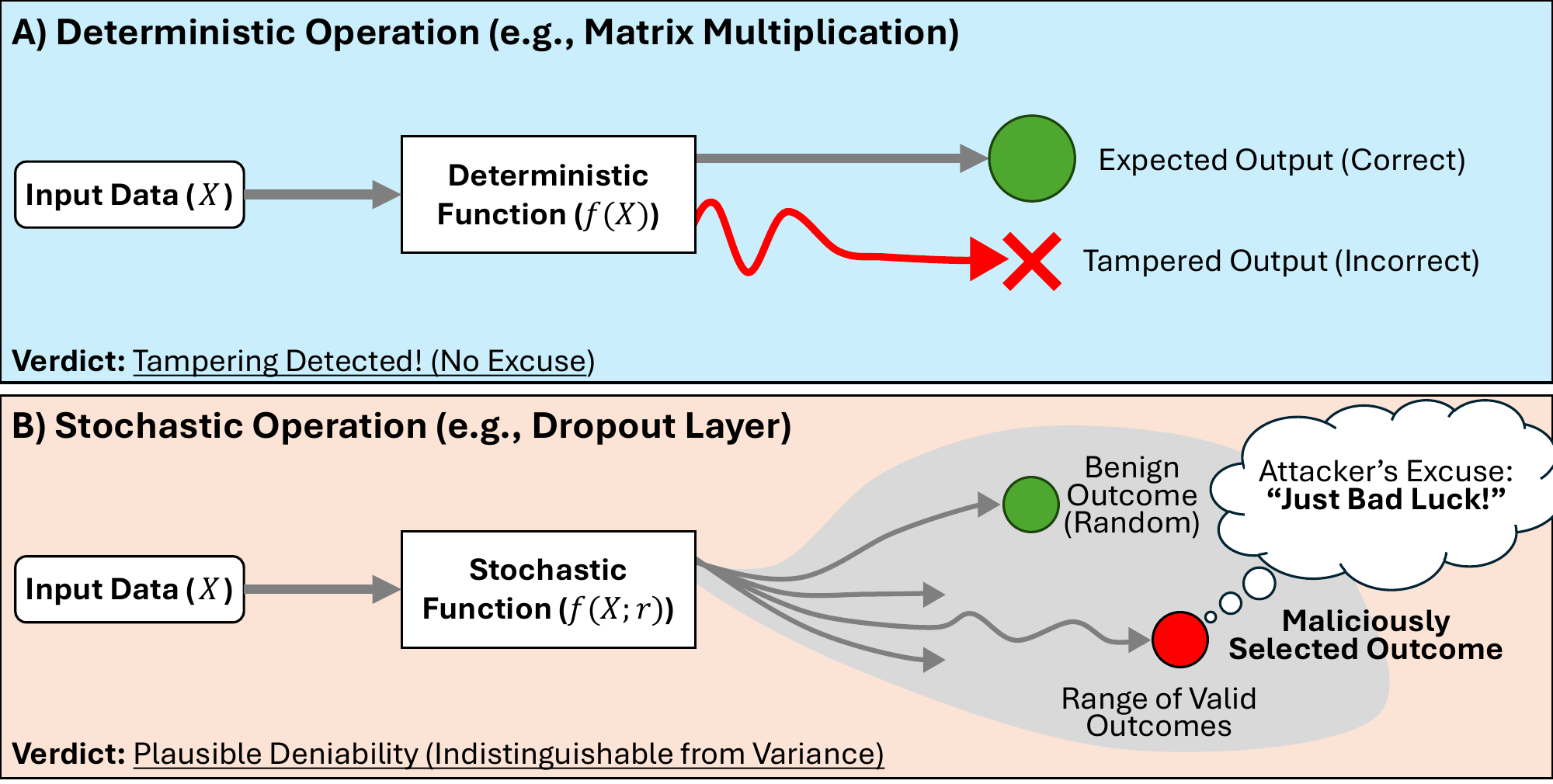}
    \vspace{-2ex}
    \caption{Deterministic vs.\ stochastic computation and the ``plausible deniability'' gap.}
    \vspace{-4ex}
    \label{fig:teaser}
\end{figure}

However, accountability is complicated by the fact that many core training steps are intentionally non-deterministic.
Stochasticity improves optimization and generalization, and common pipelines rely on randomized data shuffling, augmentation, and stochastic layer behavior.
This randomness creates an ambiguity surface: even if an adversary manipulates part of the computation, the resulting deviations can be plausibly attributed to expected run-to-run variance.
As Figure~\ref{fig:teaser} illustrates, deterministic computation enables direct comparison across executions, whereas stochastic computation introduces a ``plausible deniability'' gap.
As a result, when a user observes an unexpected outcome, logs alone often cannot answer a key question: \textit{``Did the randomness come from the intended source and was it applied correctly, or adversarially selected or altered?''}

This gap is amplified in cloud supply chains, where the training requester has limited visibility into the execution environment.
A model owner may receive only a checkpoint and logs, yet verifying that every stochastic operation was sampled honestly and applied as specified is typically infeasible.
A malicious provider or compromised worker can bias stochastic operations to degrade accuracy or implant targeted behavior while producing logs that appear benign.
Without a verifiable notion of ``correct randomness,'' such manipulation is hard to attribute or prove.

We argue that stochastic operations should be verifiable to make outsourced training trustworthy.
Concretely, a user should be able to validate, after the fact and with minimal trust assumptions, that (i) random values were generated according to an agreed-upon distribution and seed-derivation procedure, and (ii) those values were applied at the claimed steps.
We study dropout as a representative, high-impact stochastic operation.
Dropout is widely used, often enabled by default, and its mask selection directly shapes the computation graph and optimization trajectory.
If an adversary can tamper with dropout masks, they can induce systematic degradation or targeted behavior that is difficult to distinguish from ordinary variance~\cite{yuan2024dropout}.

To address this problem, we introduce \textbf{\textit{verifiable dropout}}, a mechanism that combines a verifiable random function (VRF) with succinct proofs of correct execution.
The VRF binds dropout randomness to a verifiable seed commitment, and the proof attests that the executor generated and applied the corresponding masks as claimed, without revealing the masks or other sensitive training information.
This turns \textit{``randomness as an excuse''} into \textit{``randomness as a verifiable claim''}, strengthening post hoc accountability for stochastic training in cloud ML pipelines.

Our contributions are threefold.
First, we formalize integrity risks induced by stochastic operations in cloud training supply chains and identify dropout as a concrete, security-relevant instantiation.
Second, we design a VRF- and proof-based protocol that proves correct dropout mask generation and application.
Third, we implement a prototype and evaluate its overhead in a modern training workflow, characterizing the trade-offs among proof generation cost, verification latency, and deployment constraints.

By making stochastic operations verifiable rather than merely observable, this work takes a step toward accountable training pipelines where users can audit the integrity of outsourced computation despite unavoidable randomness.

\section{Related Work}
\label{sec:relwork}
\subsection{Attacks on Stochastic Training Components}
Recent work shows that stochastic elements in training pipelines are not merely benign sources of variance, but can become an attack surface in outsourced or supply-chain settings with an untrusted executor.
Beyond classical poisoning and backdoor attacks that modify data or labels~\cite{lee2024tazza,lee2024detrigger}, attackers can manipulate randomized choices that are difficult to audit post hoc.
For example, data ordering attacks show that reordering minibatches can slow convergence or induce attacker-chosen behaviors without changing the dataset or model architecture~\cite{shumailov2021manipulating}.
More broadly, manipulating procedures that are expected to be stochastic complicates attribution because deviations can be dismissed as ordinary run-to-run variation~\cite{dahiya2024machine}.

Most closely related, \citet{yuan2024dropout} proposes Dropout Attacks, where an adversary controls which neurons are dropped instead of sampling masks uniformly at random.
They observe that typical integrity checks focus on externally visible artifacts and largely ignore verifying non-determinism, enabling plausible deniability for mask tampering.
This motivates our focus on dropout as a representative stochastic operator and our emphasis on \textit{verifiability}: rather than detecting anomalous outcomes, we enable the executor to later prove that dropout randomness was generated as specified and that the corresponding operation was applied as claimed.

\subsection{Verifiable Random Functions}
Verifiable Random Functions (VRFs) generate pseudorandom outputs together with publicly verifiable proofs of correctness, while remaining unpredictable without the secret key~\cite{micali1999verifiable}.
VRFs are widely used for trust-minimized randomness generation in distributed systems because they allow parties to validate sampled randomness without learning the secret key.

In our setting, VRFs can make the \textit{source} of dropout randomness auditable by binding randomness to agreed-upon context (e.g., step index, layer identifier, job nonce)~\cite{micali1999verifiable}.
However, VRFs alone do not certify that the sampled randomness was actually used in the training computation.
An executor can present a valid VRF proof while silently applying a different mask, motivating an additional mechanism to prove correct application.

\subsection{Proofs for Verifiable Computation and ML}
Succinct proof systems, including zero-knowledge variants, enable a prover to convince a verifier that a computation was executed correctly with low verification cost and, when needed, without revealing private inputs or intermediate states~\cite{parno2016pinocchio}.
A growing body of work on verifiable machine learning (ZKML) explores proving properties of ML inference and training while keeping data and models confidential~\cite{peng2025survey,lee2025mind}.

Our work targets a distinct verification objective: \textit{stochastic operators}.
While many ZKML efforts emphasize deterministic inference, accuracy claims, or simplified training proofs~\cite{peng2025survey}, dropout attacks arise because auditors cannot distinguish adversarial non-determinism from legitimate randomness~\cite{yuan2024dropout}.
We therefore combine VRFs with succinct proofs to close this gap, proving that the dropout mask derived from verifiable randomness was applied to the claimed tensors at the claimed steps, without exposing the mask or intermediate activations~\cite{parno2016pinocchio}.

\section{System Design}
\label{sec:system}

This section presents the design of our \textbf{Verifiable Dropout (VDO)}, which enables a verifier to check that a trainer applied \textit{dropout correctly and deterministically} according to a publicly verifiable seed, while keeping the full activation tensor private.
Our current prototype targets a practical integration path: a PyTorch training stack (Python) augmented with a zkVM-based proof backend (Rust, RISC Zero).
You can find the implementation \href{https://anonymous.4open.science/r/zkp-dropout-B79C}{\textcolor{magenta}{\underline{here}}}.
\subsection{Design Goals and Threat Model}
\paragraph{Goals.}
VDO is designed with four goals:
\begin{enumerate}[leftmargin=1.2em]
    \item \textbf{Deterministic dropout across platforms:} the dropout mask must be derived from a platform-stable PRG, independent of framework RNG differences.
    \item \textbf{Public verifiability of randomness:} the verifier should be able to validate that the dropout seed was derived from an authenticated, deterministic process.
    \item \textbf{Correctness evidence with minimal leakage:} the verifier should check correctness without receiving the full activation tensor.
    \item \textbf{Training integration:} the mechanism should attach to a standard forward pass with minimal code changes.
\end{enumerate}

\vspace{-0.3em}
\paragraph{Threat model.}
We consider a \textbf{potentially dishonest trainer} who may claim to have applied dropout but instead uses a different mask, different dropout probability, or an adversarial transformation (i.e., manipulating the mask for degrading accuracy).
The trainer controls the training environment and software supply chain, so verifiers cannot blindly trust local logs.
The verifier is assumed to know public metadata (model identifier, training step, batch id) and supplies a fresh nonce to prevent precomputation.
\begin{figure}
    \centering
    \includegraphics[width=1\linewidth]{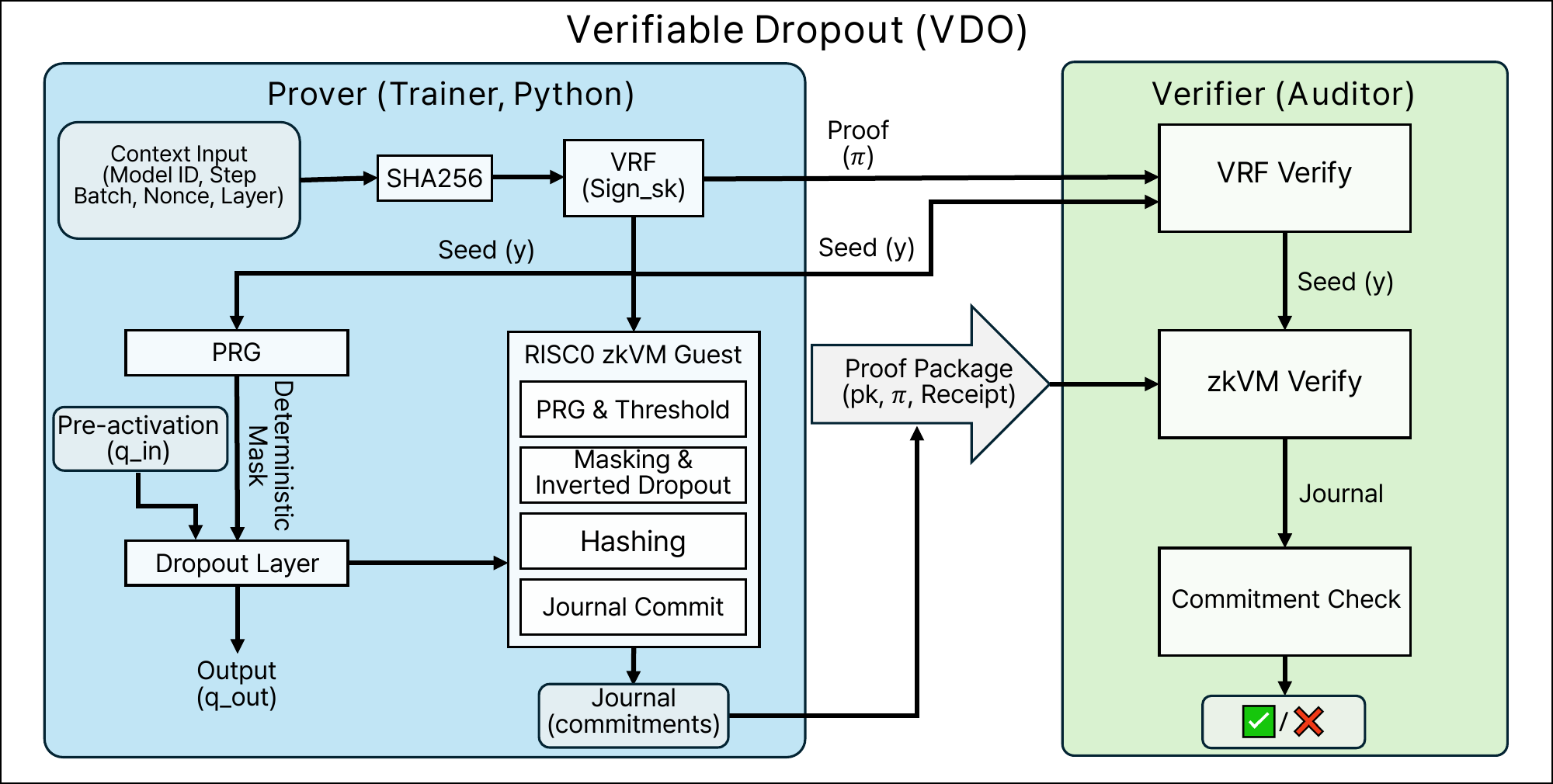}
    \vspace{-2ex}
    \caption{Overview workflow of VDO.}
    \vspace{-4ex}
    \label{fig:overview}
\end{figure}
\subsection{System Overview}
Figure~\ref{fig:overview} (conceptual) summarizes our architecture. VDO spans three layers:

\begin{enumerate}[leftmargin=1.2em]
    \item \textbf{Python training wrapper}: provides a \texttt{VerifiableDropout} module compatible with \texttt{nn.Module} and a \texttt{VerifiableRunner} that executes a forward pass while collecting proof artifacts.
    \item \textbf{zkVM guest}: implements a deterministic dropout transformation over quantized activations and commits a compact journal containing hashes.
    \item \textbf{Host}: produces a receipt/journal from JSON input, verifies the receipt, and extracts journal fields for Python-side checking.
\end{enumerate}

\subsection{Verifiable Random Seed Derivation}
At the core of VDO is a \emph{publicly verifiable} and \emph{deterministic} seed that binds dropout randomness to the training context. We use a VRF-like construction based on deterministic Ed25519 signatures.

\vspace{-0.3em}
\paragraph{Context binding.}
For each dropout layer $\ell$, step $t$, and batch id $b$, the trainer forms a context:$ \textsf{ctx} = \textsf{pack}( \textsf{model\_id}, t, b, \textsf{nonce}, \ell )$
and hashes it to obtain the VRF input $x = \textsf{SHA256}(\textsf{ctx})$.

\vspace{-0.3em}
\paragraph{Evaluation.}
Given a signing key $\textsf{sk}$, the trainer computes a deterministic signature $\pi = \textsf{Sign}_{\textsf{sk}}(x)$ and sets the dropout seed as $y = \textsf{SHA256}(\pi)$ (32 bytes). The proof packet includes $(\textsf{pk}, x, y, \pi)$; the verifier checks $\textsf{Verify}_{\textsf{pk}}(x,\pi)$ and recomputes $y$.

\vspace{-0.3em}
\paragraph{Why signatures.}
While this is not a standardized VRF construction, it provides the two properties we need in practice: (i) determinism and unpredictability of $y$ without $\textsf{sk}$, and (ii) public verifiability via $\textsf{pk}$. This makes it a convenient building block for verifiable dropout randomness.

\subsection{Platform-Stable PRG and Mask Generation}
To avoid differences in RNG implementations across platforms, VDO derives the dropout mask using a hash-expander PRG.

\vspace{-0.3em}
\paragraph{PRG definition.}
We expand the 32-byte seed $y$ into a stream of bytes using:
$\textsf{block}_i = \textsf{SHA256}\big( y \,\|\, \textsf{ctr}_{i}^{\textsf{le64}} \big)$
concatenating blocks until enough bytes are produced. We interpret each block as eight \texttt{u32} values using \textbf{little-endian} decoding. This PRG is implemented identically in Python and the zkVM guest.

\vspace{-0.3em}
\paragraph{Thresholding.}
We represent dropout probability as an exact rational $\frac{p_{\textsf{num}}}{p_{\textsf{den}}}$ to avoid floating-point drift. For each element, we sample $u \in [0,2^{32})$ and keep the element if $u \ge \lfloor p \cdot 2^{32} \rfloor$. The resulting keep mask is a byte vector.

\subsection{Correctness Evidence via zkVM}
VDO generates a zkVM receipt to attest that the dropout transformation adheres to the specified probability parameters ($p_{\textsf{num}}, p_{\textsf{den}}$), the derived seed $y$, and the pre-dropout activations. This mechanism guarantees the correctness of the stochastic operation without disclosing the full activation tensor to the verifier.

\vspace{-0.3em}
\paragraph{Quantization and Integer Arithmetic.}
To ensure computational efficiency and deterministic verification within the zkVM guest, we quantize the pre-dropout activation tensor $x$ into a flattened integer vector $q = \textsf{round}(x \cdot S) \in \mathbb{Z}^{n}$, where $S$ is a fixed scaling factor (e.g., 65536). The guest performs the dropout operation on this quantized vector using integer arithmetic. For elements retained by the mask, the guest computes the scaled value as
$q' = \textsf{round}\left( q \cdot \frac{p_{\textsf{den}}}{p_{\textsf{den}} - p_{\textsf{num}}} \right),$
while outputting $0$ for dropped elements. We implement \emph{round-to-nearest} signed division and clamp the results to integer boundaries. Crucially, to ensure consistency with the host verification, the guest serializes the output vector as little-endian bytes prior to hashing.

\vspace{-0.3em}
\paragraph{Committed Journal.}
The zkVM guest commits to a compact cryptographic journal comprising three key elements: the SHA256 hash of the generated mask bytes, the SHA256 hash of the serialized output vector, and the total number of elements ($n$). Upon completion, the host retrieves the receipt and verifies that the committed journal matches the metadata recorded in the proof object generated during training.

\subsection{End-to-End Protocol}
The execution workflow for a verifiable dropout layer integrates cryptographic proof generation with the standard training loop. First, the system derives the verifiable seed components $(x,\pi,y)$ from the training context and a cryptographic signature. Following this, the standard floating-point forward pass derives the dropout mask using the PRG and applies inverted dropout to produce the output used by subsequent model layers.

To generate the proof, the system quantizes the input $x$ into $q$ and invokes the zkVM prover with the derived seed, probability parameters, and the quantized input. This process yields a zkVM receipt and journal hashes, which are then packaged into a consolidated proof object alongside the VRF packet.

\vspace{-0.3em}
\paragraph{Verification.}
The verification process involves a tripartite check to ensure integrity. The verifier first validates the VRF proof by checking the signature $\pi$ with the public key $\textsf{pk}$ and confirming the seed derivation. It then validates the zkVM receipt via the verification interface and finally ensures that the journal committed within the receipt is consistent with the claimed statement fields in the proof object.
% \subsection{Implementation Notes and Current Limitations}
% \paragraph{Dev mode.}
% RISC Zero supports a development mode (\texttt{RISC0\_DEV\_MODE=1}) that accelerates proving but does not generate production-secure receipts. Our experiments explicitly record whether dev mode is enabled.

% \paragraph{What is verified today.}
% The current prototype proves the \emph{internal consistency} of dropout with respect to $(y,p,q)$ and commits to hashes of the quantized output and mask.

% \paragraph{Open gap.}
% End-to-end verification of the \emph{full} forward pass (i.e., that $q$ genuinely corresponds to the model's true pre-dropout activations) requires extending the statement to bind:
% (1) the model parameters (or their commitment),
% (2) the input activation hash/commitment, and
% (3) the linkage between float activations and quantized vector $q$.
% We view this as a natural next step, and the current design cleanly isolates the dropout subroutine as a verifiable building block.

% Optional: put a figure placeholder
% \begin{figure}[t]
%   \centering
%   \includegraphics[width=0.95\linewidth]{figures/vdo_overview.pdf}
%   \caption{System overview of Verifiable Dropout (VDO).}
%   \label{fig:overview}
% \end{figure}

\section{Evaluation}
\label{sec:eval}

We evaluate VDO along four axes: (i) \textbf{cross-implementation determinism} between the local reference and the RISC~Zero zkVM guest, (ii) \textbf{end-to-end proving overhead} of the zkVM-backed dropout layer, (iii) \textbf{scalability} with the number of dropout elements, and (iv) \textbf{tamper detection} against representative manipulation attempts on stochastic layers.

\vspace{-0.3em}
\paragraph{Setup.}
Unless otherwise stated, we generate a per-layer seed by signing a context string that binds \texttt{model\_id}, training \texttt{step}, \texttt{batch\_id}, verifier-provided \texttt{nonce}, and \texttt{layer\_id}. We derive a deterministic dropout mask from the seed using a hash-expander PRG and apply inverted dropout. The prover produces a zkVM receipt whose journal commits to the dropout mask hash and the quantized post-dropout output hash; the verifier validates the VRF packet and verifies the receipt.

\begin{figure}[t]
    \centering
    \begin{minipage}[t]{0.475\columnwidth}
        \centering
        \includegraphics[width=0.99\linewidth]{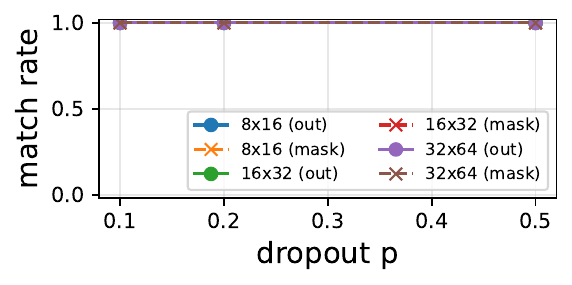}
        \vspace{-4ex}
        \caption{Local reference vs. zkVM commitments across shapes and dropout $p$.}
        \label{fig:exp1-1}
    \end{minipage}
    \hfill
    \begin{minipage}[t]{0.475\columnwidth}
        \centering
        \includegraphics[width=0.99\linewidth]{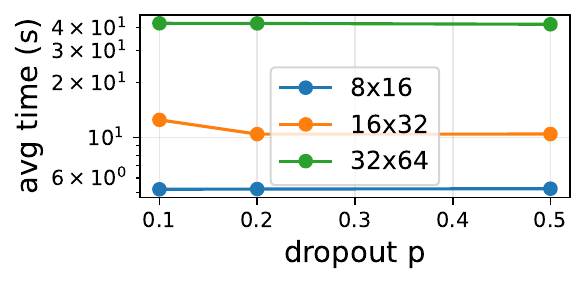}
        \vspace{-4ex}
        \caption{Average runtime vs. dropout $p$ for each shape.}
        \label{fig:exp1-2}
    \end{minipage}
\end{figure}

\subsection{Exp1: Cross-Implementation Determinism}
To remove plausible deniability, VDO must ensure that the dropout randomness is deterministic and that both sides can derive the same statement to verify. We therefore compare the Python reference and zkVM guest under identical inputs, checking whether they agree on the dropout mask commitment and the quantized post-dropout output commitment. Across tensor shapes and dropout probabilities, the two implementations consistently produce matching mask commitments (Figure~\ref{fig:exp1-1}), indicating that seed expansion, thresholding, and serialization are aligned. Once the quantization and inverted-dropout scaling conventions are matched, the output commitments also agree, confirming end-to-end determinism of the proved computation. Figure~\ref{fig:exp1-2} reports the latency of VDO runtime primarily tracks tensor size and is comparatively insensitive to $p$, consistent with proof generation being dominated by linear-time hashing and commitment computation over the activation footprint.

\subsection{Exp2: Runtime Overhead in a Forward Pass}
We next quantify the overhead of VDO in a training-like forward path by comparing three variants: the framework's Baseline dropout, a Hash-only variant that computes deterministic masks and local commitments without producing a receipt, and VDO that generates a zkVM receipt. Figure~\ref{fig:exp2} shows that Hash-only adds modest overhead relative to Baseline, isolating the cost of deterministic mask generation and hashing, while VDO incurs substantially higher latency due to zkVM proving. This separation clarifies that the dominant cost comes from proof generation rather than from making dropout deterministic, and suggests that practical deployments should amortize proving by sampling layers/steps or emitting receipts at audit checkpoints rather than proving every invocation~\cite{boo2021litezkp}.

\subsection{Exp3: Scalability with Number of Elements}
\begin{figure}[t]
    \centering
    \begin{minipage}[t]{0.475\columnwidth}
        \centering
        \includegraphics[width=1.\linewidth]{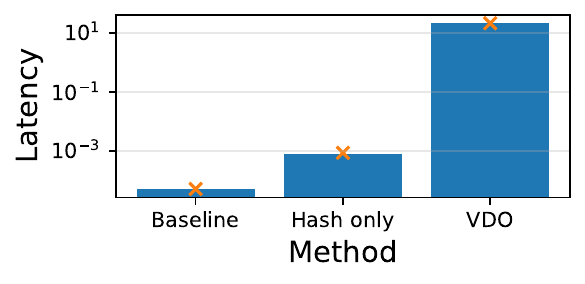}
        \vspace{-4ex}
        \caption{Per-forward time for Baseline dropout, Hash-only, and VDO.}
        \label{fig:exp2}
    \end{minipage}
    \hfill
    \begin{minipage}[t]{0.475\columnwidth}
        \centering
        \includegraphics[width=1.\linewidth]{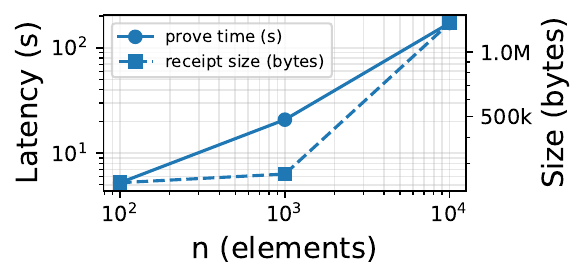}
        \vspace{-4ex}
        \caption{Proving latency and receipt size vs. number of elements $n$.}
        \label{fig:exp3}
    \end{minipage}
\end{figure}

To understand the scaling behavior, we vary the number of dropout elements $n$ (flattened activation length) and measure end-to-end proving latency and receipt size. As shown in Figure~\ref{fig:exp3}, both proving latency and receipt size grow monotonically with $n$, reflecting the linear dependence on the amount of data that must be committed and processed inside the zkVM. These results highlight a fundamental constraint of per-layer, per-step enforcement for large activations and further motivate selective proving strategies when applying VDO to larger models.

\subsection{Exp4: Attack Detection}
Finally, we validate that VDO catches representative manipulations that exploit stochasticity in a supply-chain setting. Starting from a valid transcript, we attempt (i) replacing the VRF-derived seed, (ii) claiming a different dropout probability $p$ than what was proved, and (iii) modifying the claimed quantized activation while reusing the receipt. In all cases, verification fails because the VRF packet and receipt jointly bind the statement to the specific context, dropout configuration, and committed post-dropout output. Table~\ref{tab:tamper_detect_rate} summarizes the results: all three attacks are detected in our test suite.

\begin{table}[t]
\centering
\begin{adjustbox}{width=1.\linewidth}
\begin{tabular}{lccc}
\toprule
 & Seed tamper & $p$ tamper & Activation tamper \\
\midrule
Detection Rate & 100\% & 100\% & 100\% \\
\bottomrule
\end{tabular}
\end{adjustbox}
\caption{Detection rate for seed, $p$, and activation tampering.}
\label{tab:tamper_detect_rate}
\end{table}

\section{Limitation and Future Work}
\label{sec:limit}
While this work successfully bridges the ``plausible deniability" gap by establishing a rigorous framework for verifying stochastic operations, the current computational overhead of generating zero-knowledge proofs poses a challenge for large-scale deployment. Our evaluation demonstrates that while the security guarantees are solid, the latency introduced by the general-purpose zkVM backend can become a bottleneck in high-throughput training pipelines, particularly as the number of stochastic elements grows. This trade-off between cryptographic rigor and training efficiency remains the primary limitation preventing immediate, universal adoption in time-sensitive environments.

Future work will therefore focus on optimizing this balance to transition Verifiable Dropout from a robust theoretical mechanism to a practical, everyday tool. We aim to investigate probabilistic auditing strategies, where only a random subset of stochastic layers is verified to amortize proving costs without compromising the deterrent effect against attackers. Additionally, exploring specialized arithmetic circuits for tensor operations and hardware-accelerated proving backends offers a promising path to significantly reduce latency. These advancements are essential to deliver a seamless, low-overhead solution that ensures accountability in modern, large-scale AI supply chains.
\section{Conclusion}
\label{sec:conclusion}
This work addresses the critical integrity gap in outsourced AI training, where stochastic operations have historically provided attackers with plausible deniability. By introducing Verifiable Dropout, we successfully transform randomness from an unverifiable excuse into a cryptographically binding claim, enabling the post-hoc auditing of dropout masks without compromising model confidentiality. Although computational trade-offs remain for large-scale deployment, our framework establishes a fundamental precedent for trustworthy AI supply chains, demonstrating that even non-deterministic computation can be made strictly accountable.
% \begin{acks}
% ACK!
% \end{acks}

%% ack stuff
%\begin{acks}
%\end{acks}

\balance
\bibliographystyle{ACM-Reference-Format}
\bibliography{reference}

\clearpage
%%%%%%%%%%%%%%%%%%%%%%%%%%%%%%%%%%%%%%%%%%%%%%%%%%%%%%%%%%%%%%%%%%%%%%%%%%%%%%%%%%%%

\end{document}